\documentclass[]{spie}

\usepackage[utf8]{inputenc}

\usepackage{tikz}
\usepackage{pgfplots}
\pgfplotsset{compat=1.5}

\usepackage{amsmath,amsfonts,amssymb}

\usepackage{graphicx}
\usepackage[colorlinks=true, allcolors=blue]{hyperref}

\usepackage{txfonts}
\usepackage{textcomp}
\usepackage{eurosym}

\newcommand{\micron}{\mbox{$\mu$m}}
\renewcommand{\deg}{\mbox{deg}}
\newcommand{\sqdeg}{\mbox{$\deg^2$}}

\newcommand{\arcsec}{\mbox{arcsec}}

\newcommand{\persqarcsec}{\mbox{\arcsec$^{-2}$}}

\title{DDOTI: the deca-degree optical transient imager}

\author[a]{Alan~M.~Watson}
\author[a]{William~H.~Lee}
\author[b,c]{Eleonora~Troja}
\author[d]{Carlos~G.~Román-Zúñiga}
\author[e]{Nathaniel~R.~Butler}
\author[b,f]{Alexander~S.~Kutyrev}
\author[b]{Neil~A.~Gehrels}
\author[a]{Fernando~Ángeles}
\author[g]{Stéphane~Basa}
\author[h]{Pierre-Eric~Blanc}
\author[i]{Michel~Boër}
\author[a]{Jose~A.~de~Diego}
\author[a]{Alejandro~S.~Farah}
\author[d]{Liliana~Figueroa}
\author[a]{Yilen~Gómez~Maqueo~Chew}
\author[j]{Alain~Klotz}
\author[d]{Fernando~Quirós}
\author[d]{Maurico~Reyes-Ruíz}
\author[a]{Jaime~Ruíz-Diáz-Soto}
\author[k]{Pierre~Thierry}
\author[a]{Silvio~Tinoco}

\affil[a]{Instituto de Astronomía, Universidad Nacional Autónoma de México, Apartado Postal 70-264, 04510~México, México}
\affil[b]{NASA, Goddard Space Flight Center, Greenbelt, MD 20771, USA}
\affil[c]{CRESST, University of Maryland, College Park, MD, USA}
\affil[d]{Instituto de Astronomía, Universidad Nacional Autónoma de México, Apartado Postal 106, 22860~Ensenada, Baja California, México}
\affil[e]{School of Earth and Space Exploration, Arizona State University, Tempe, AZ 85287, USA}
\affil[f]{Astronomy Department, University of Maryland, College Park, MD, USA}
\affil[g]{Aix Marseille Université, CNRS, LAM (Laboratoire d’Astrophysique de Marseille) UMR 7326, 13388 Marseille, France}
\affil[h]{Observatoire de Haute-Provence, 04870 Saint-Michel l’Observatoire, France}
\affil[i]{ARTEMIS, UMR 7250 (CNRS/OCA/UNS), Boulevard de l'Observatoire
CS 34229, F-06304 Nice Cedex 4, France}
\affil[j]{CESR, Observatoire Midi-Pyrénées, CNRS, Université de Toulouse, BP 44346, F-31028 Toulouse Cedex 4, France}
\affil[k]{Observatoire de Roudiere, 1732 Chemin des Crêtes, 3190 Auragne, France}

\authorinfo{Further author information: Send correspondence to A.M.W. E-mail: alan@astro.unam.mx}

\begin{document} 
\maketitle

\begin{abstract}
DDOTI will be a wide-field robotic imager consisting of six 28-cm telescopes with prime focus CCDs mounted on a common equatorial mount. Each telescope will have a field of view of 12 {\sqdeg}, will have 2 arcsec pixels, and will reach a $10\sigma$ limiting magnitude in 60 seconds of $r \approx 18.7$ in dark time and $r \approx 18.0$ in bright time. The set of six will provide an instantaneous field of view of about 72 {\sqdeg}. DDOTI uses commercial components almost entirely. The first DDOTI will be installed at the Observatorio Astronómico Nacional in Sierra San Pedro Martír, Baja California, México in early 2017. The main science goals of DDOTI are the localization of the optical transients associated with GRBs detected by the GBM instrument on the Fermi satellite and with gravitational-wave transients. DDOTI will also be used for studies of AGN and YSO variability and to determine the occurrence of hot Jupiters. The principal advantage of DDOTI compared to other similar projects is cost: a single DDOTI installation costs only about US\$500,000. This makes it possible to contemplate a global network of DDOTI installations. Such geographic diversity would give earlier access and a higher localization rate. We are actively exploring this option.
\end{abstract}

\keywords{wide-field imagers, robotic telescopes, optical transients, gamma-ray bursts, gravitational-wave transients, synoptic observations}

\section{INTRODUCTION}

DDOTI is a wide-field imager employing six 28-cm telescopes with prime focus CCDs mounted on a common equatorial mount. Each telescope will have a field of view of 12 {\sqdeg} with 2 arcsec pixels, and the set of six will provide an instantaneous field of view of about 72 {\sqdeg}. DDOTI uses commercial components almost entirely. The only custom components are the mechanical adapters for the telescopes and the CCDs and the electronics cabinets. The first DDOTI will be installed at the Observatorio Astronómico Nacional (OAN) in Sierra San Pedro Martír (SPM), Baja California, México in early 2017. The $10\sigma$ limiting magnitude in 60 seconds will be $r \approx 18.7$ in dark time and $r \approx 18.0$ in bright time. 

DDOTI is specifically designed for the localization of the optical transients associated with gamma-ray bursts (GRBs) detected by the gamma-ray burst monitor (GBM) instrument on the Fermi satellite and with gravitational-wave transients. It will also be used for synoptic studies.

The principal advantage of DDOTI compared to other similar projects such as iPTF, ZTF, and BlackGEM is cost: a single DDOTI installation costs only about US\$500,000. This makes it possible to contemplate a global network of DDOTI installations dedicated to localizing the optical counterparts of Fermi/GBM GRBs and gravitational-wave transients. Such geographic diversity would give earlier access and a higher localization rate.

In this contribution, we will discuss our reference science cases, the DDOTI hardware and software, our tests on the sky of the single-telescope prototype, the state of our six-telescope imager for OAN/SPM, and our plans for a future network.

\section{REFERENCE SCIENCE CASES}

\subsection{Localization of Fermi/GBM Gamma Ray Bursts}

GRBs are the brightest explosions in the universe. GRBs can be divided into three classes according to the time during which they emit gamma rays and these classes are believed to correspond to different progenitors and explosion mechanisms\cite{kouveliotou-1993,woosley-2006,lee-2007,piro-2014}. Short GRBs emit for up to about 2 seconds and are thought to arise in the merger of two compact objects (neutron stars or black holes). Long GRBs emit from about 2 second to about 1000 seconds and are thought to arise from the core-collapse of a massive star. Ultra-long GRBs emit for 10,000 seconds or longer, and are not well understood but might arise from core-collapse in blue supergiants. GRBs are interesting as laboratories for the physics of ultra-relativistic plasmas (with Lorentz factors of 100 to 1000), from the standpoint of stellar astrophysics, as light-bulbs to study the intergalactic and interstellar medium at high redshift well beyond the epoch of the quasars, and as tracers of star formation.

GRBs are detected by a number of satellites, but NASA’s Swift and Fermi satellites provide the largest number of detections. To fully exploit these detections, we need imaging and spectroscopic observations of the optical/infrared counterpart to measure the light curve, the redshift, and identify the host galaxies. The first step toward this is a precise localizations of the counterpart to within a few arcsec.

The Swift satellite detects about 90 GRBs each year with its wide-field Burst Alert Telescope (BAT). By itself, BAT localizes the GRBs with a precision of about 3 arcmin. However, Swift responds to most of these detections by automatically repointing to observe with its narrow-field X-Ray Telescope (XRT), which localizes the GRB and/or its afterglow with a precision of about 2 arcsec. These localizations are distributed over the Internet typically within 3 minutes of the burst. In many cases, subsequent observations with ground-based telescopes such as GROND\cite{greiner-2008} and RATIR\cite{butler-2012} yield a localization of the counterpart to within an arcsec. This chain of localizations eventually leads to redshifts for about one third of Swift/BAT GRBs.

The scenario for Fermi/GBM GRBs is quite different. Its wide-field GBM gamma-ray instrument detects about 250 GRBs each year and gives compact localizations typically of order 100 {\sqdeg}. Compared to Swift/BAT GRBs, these GRBs are interesting for three reasons. First, the detection rate is 2.5 times higher. Second, the spectral response of Fermi/GBM extends to higher energy than that of Swift/BAT and so can see the peak of the gamma-ray spectrum, which is important for understanding the burst. Finally, Fermi/GBM detects about 45 short GRBs per year whereas Swift/BAT detects only about 9 per year. This is because Fermi/GBM detects more GRBs and because short GRBs tend to be harder and so are missed by the softer response of Swift/BAT. The full scientific exploitation of a Fermi/GBM GRB requires a localization to within a few arcsec. At the moment, there are three routes to improving the initial GBM localizations:

\begin{enumerate}

\item
About 10\% of GBM GRBs are also detected by Swift/BAT\cite{stamatikos-2009}, which allows further observations with the well-developed chain of lower-energy instruments described above.

\item
About 7\% of GBM GRBs are also detected by Fermi Large Area Telescope (LAT) instrument with localizations of typically 10 arcmin. For about 2/3 of these, the localizations are improved by target-of-opportunity observations with Swift/XRT and narrow-field ground-based instruments. However, for about 1/3 of the events are not improved. (The Fermi/LAT GRB catalog has 83 bursts in 2009--2014 and for 29 of these the best localization is by LAT.)

\item
A small number of GBM GRBs are followed-up by with the iPTF\cite{singer-2015}, a sensitive wide-field imager with a field of 7 {\sqdeg} and a $10\sigma$ limiting magnitude of $R \approx 19.4$ in 60 seconds. In 35 attempts with delays between 6 minutes and 24 hours, iPTF has managed to detect 8 counterpart afterglows associated with Fermi/GBM GRBs\cite{singer-2015}. These eight detections occurred over a period of 13 months, and so represent only about 3\% of the Fermi/GBM bursts.
\end{enumerate}

DDOTI will carry forward the pioneering work carried out by iPTF. It will have less aperture, but a significantly larger field of view. Our modelling suggests that it will be able to localize about 45 GBM GRBs per year. This will double the number of precise localizations from about 20\% to about 40\% and will significantly advance our ability to effectively study these GRBs.

DDOTI will attempt to follow-up \emph{every} Fermi/GBM GRB accessible from SPM ($\delta > -30$ \deg). The only higher-priority science with DDOTI will be searches for counterparts of gravitational-wave transients. We anticipate using about 40\% of the available observing time for this program.

\subsection{Counterparts of Gravitational-Wave Transients}

When we conceived DDOTI in the early summer of 2015, we proposed to use it to search for the optical counterparts of gravitational-wave transients detected by Advanced LIGO and Virgo. This interest has been strengthened by the astonishing early detection of the gravitational-wave transient GW150914 in the first Advance LIGO observing run\cite{Abbott:2016bn}. The error regions for gravitational-wave transients are currently typically 600 {\sqdeg}, which is even larger than the uncertainty for Fermi/GBM detections, but are expected to drop to around 100 {\sqdeg} as other observatories are commissioned. Our observations will complement the efforts of other teams with narrow-field instruments, who typically target nearby galaxies within the uncertainty region. We anticipate using about 10\% of the available observing time for this program.


\subsection{Other Synoptic Science}

Searches for counterparts of Fermi/GBM GRBs and gravitational-wave transients are our top priorities for DDOTI. However, we expect them to occupy only about 50\% of the total time. We will fill the remaining time with three science projects:

\begin{enumerate}
\item A study to look for new AGNs through their correlated temporal variability \cite{butler-2011}, which avoids the problems with using color-color diagrams above high-redshift or reddened AGN\cite{richards-2009}. We anticipate using about 25\% of the available time for this program.

\item A study to characterize the variability of known YSOs and to look for new YSOs through their variability by mapping synoptically entire areas of a collection of star forming regions within 2~kpc. This search will avoid the known problems of using color-color and color-magnitude diagrams\cite{briceno-2005}. Our work will extend the earlier studies\cite{pojmanski-2002,wozniak-2004,sesar-2011} to greater sensitivity, larger sky coverage, and better temporal sampling. We anticipate using about 15\% of the available time for this program.

\item A study to measure the fraction of hot Jupiters. Kepler found\cite{howard-2012} a lower frequency of hot Jupiters (0.1\%) compared to estimates from the ground\cite{wright-2012} (about 1\%). The wide field of DDOTI will allow us to look for hot Jupiters in a much larger sample of solar-type stars and resolve this discrepancy. We anticipate using about 10\% of the available time for this program.
\end{enumerate}

\section{Hardware and Software}

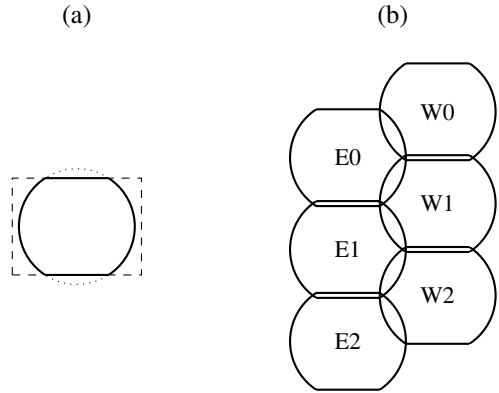
\begin{figure}
\begin{center}
\newcommand{\fov}[1]{
  \begin{scope}
    \clip (-24.528,-18.396) rectangle (+24.528,+18.396);
    \draw[thick] (0,0) circle (22) node{\footnotesize #1};
 \end{scope}
 \draw[thick] ({-sqrt(22*22-18.396*18.396)},-18.396) -- ({+sqrt(22*22-18.396*18.396)},-18.396);
 \draw[thick] ({-sqrt(22*22-18.396*18.396)},+18.396) -- ({+sqrt(22*22-18.396*18.396)},+18.396);
}
\newcommand{\boresight}{
  \draw (-10,0) -- (+10,0);
  \draw (0,-10) -- (0,+10);
}
\begin{tikzpicture}[scale=0.035]
\begin{scope}[xshift=0cm]
  \draw (0,80) node {(a)};
  \fov{}
  \draw[dashed] (-24.528,-18.396) rectangle (+24.528,+18.396);
  \draw[dotted] (0,0) circle (22);
\end{scope}
\begin{scope}[xshift=120cm]
 \draw (0,80) node {(b)};
 \begin{scope}[rotate=0]
  \begin{scope}[xshift={-0.5*34.066cm},yshift={1.425*18.366cm}]\fov{E0}\end{scope}
  \begin{scope}[xshift={0.5*34.066cm},yshift={-1.425*18.366cm}]\fov{W2}\end{scope}
  \begin{scope}[xshift={0.5*34.066cm},yshift={2.375*18.366cm}]\fov{W0}\end{scope}
  \begin{scope}[xshift={-0.5*34.066cm},yshift={-2.375*18.366cm}]\fov{E2}\end{scope}
  \begin{scope}[xshift={0.5*34.066cm},yshift={0.475*18.366cm}]\fov{W1}\end{scope}
  \begin{scope}[xshift={-0.5*34.066cm},yshift={-0.475*18.366cm}]\fov{E1}\end{scope}
 \end{scope}
\end{scope}
\end{tikzpicture}
\end{center}
\caption{(a): The field of a single basic imager. The FLI ML50100 CCD is $49.1 \times 36.8$ mm or $4.54 \times 3.40$ {\deg} (the dashed rectangle), but beyond a diameter of 44 mm or 4.07 {\deg} (the dotted circle) vignetting and optical aberrations become significant. The final effective field of view (the thick line) is circular with two missing circular segments and amounts to about 12 {\sqdeg}. This field can be oriented arbitrarily on the sky by rotating the CCD. (b) The combined field for the six basic imagers. This field is approximately $11 \times 7$ {\deg} or 72 {\sqdeg}.}
\label{figure:field}
\end{figure}

DDOTI consists of six “basic imagers” on a common mount. Each basic imager consists of a Celestron Rowe-Ackermann Schmidt Astrograph, a Starlight Instruments Focuser Boss II focus drive, and a Finger Lakes Instrument ML50100 CCD. The combination of astrograph and detector gives each basic imager a nominal field of about $4.5 \times 3.4$ {\deg} or 12 {\sqdeg} with 2.0 arcsec pixels. The common mount is an Astelco NTM-500 equatorial mount, which accommodates three basic imagers on each side using a custom adapter. The adaptor points each basic imager in a slightly different direction to give a total field of about $11 \times 7$ {\deg} or 72 {\sqdeg} (see Figures~\ref{figure:field} and \ref{figure:mount}).

\subsection{Astrograph}

The Celestron Rowe-Ackermann Schmidt Astrograph (RASA) is a prime-focus Schmidt astrograph with an aperture of 279 mm (11-inch), a focal length of 620 mm, a final focal ratio of $f/2.22$, and a focal plane that is 70 mm in diameter. Celestron began to sell the RASA in 2014. 

The RASA stands out from earlier ad hoc prime-focus imagers (Fastar and Hyperstar) by its much larger imaging field and higher image quality. Celestron state that the RMS spot size at field diameter of 44 mm (the diagonal of the detector of a full-frame DSLR) is about 2 {\micron} RMS from 400--700 nm. If the images are roughly Gaussian, this corresponds to roughly 1.1 arcsec FWHM, which is smaller than our pixel size. The relative illumination at the edge of this field is 78\%. Presumably the image quality decreases between 44 mm and 70 mm and redward of 700 nm, but Celestron has not published data on this.


The detector and corrector optics are supported by the Schmidt objective lens. One worry, therefore, is that the weight of the detector might cause distortions in the objective and flexions of the detector plane out of the telescope focal plane. We investigated this with the prototype imager, and our initial results suggest that that there is some flexure of the focal plane and this results in defocusing to a geometric image diameter of 2 arcsec at a field radius of 1.7 \deg. The degradation appears to be linear, so most of the field sees better image quality. This degree of defocus is acceptable for our applications.

\subsection{Focuser}

We equip each RASA with a Starlight Instruments Focuser Boss II controller with a a HSM20 motor. The motor couples to the shaft of the fine adjustment wheel of the focus mechanism of the primary mirror. The Focuser Boss II uses an Optec Focus Lynx Hub and communicates with the control computer over a USB-to-serial adaptor.

\subsection{Detector}

The Finger Lakes Instrument ML50100 packages an ON Semi KAF-50100 CCD in an extremely compact head. The CCD has a format of $8176\times6132$ with 6 {\micron} square pixels. It can be read through two channels at 8 MHz, which gives a nominal read time of 3 seconds. At this speed, the read noise is 10--15 electrons. The quantum efficiency of the detector is 40\% at 400 nm, rising to 60\% at 550 nm, and then falling again to 45\% at 700 nm. The interface to the control computer is USB2.0. The outer diameter of the head is 128.7 mm, largely dictated by the 65 mm shutter, which exceeds the central obscuration of the astrograph. The ML50100 became available for astronomical use at the end of 2015.

The detectors are equipped with Melles-Griot UltraThin 65-mm shutters which have a mean-time-to-failure of one million cycles. If we assume one exposure per minute for 10 hours per night for 4 nights out of 5, we expect the mean-time-to-failure to be about 5 years per shutter. Since we have six shutters, we expect about one failure per year.

\subsection{Detector Adapter}


We have designed and fabricated an adapter to mount the ML50100 CCD for the RASA telescope. Our first iteration did not have tip-tilt adjustment. However, in initial testing of the DDOTI prototype on sky it was obvious that the CCD was not perfectly mechanically aligned and in the fast beam this resulted in a significant focus shift over the field. Our next iteration will have static tip-tilt adjustment using three push-pull adjustment bolts.

\subsection{Filter}

The RASA is designed to work from 400-700 nm. We are interested in extending the response beyond 700 nm in order to have sensitivity to high-redshift GRBs. (The Ly-$\alpha$ forest at $z\approx4.75$ reaches 700 nm.) There are two caveats in working beyond 700 nm; one is that the RASA might show chromatic aberration and the other is that it might have low efficiency. We have acquired 400--700 and 400-800 nm filters to investigate chromatic aberration in the DDOTI prototype at SPM. We will also explicitly measure the efficiency. The results of these investigations will inform our decision of whether it is necessary to use a filter in the complete six-telescope imager.

\subsection{Equatorial Mount}

\begin{figure}
\begin{center}
\includegraphics[width=0.7\linewidth]{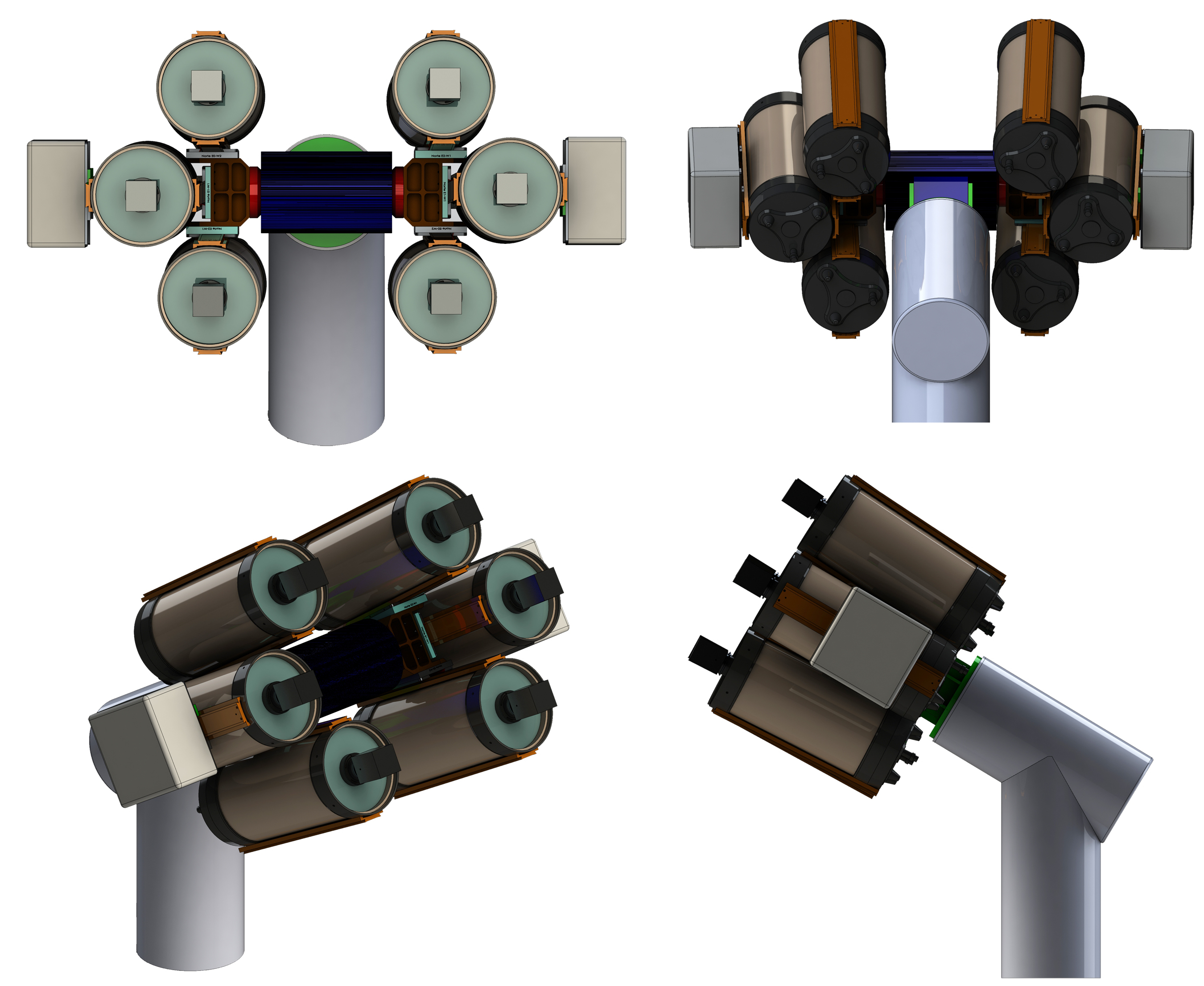}
\end{center}
\caption{The mechanical design of DDOTI. The Astelco NTM-500 mount is shown in purple and is used in an equatorial configuration. The steel pier is designed to avoid mechanical interferences with the telescopes and to permit them to point to any point in the sky. The telescopes are shown pointing towards the celestial north, but each telescope is offset slightly in angle to give the coverage indicated in Figure~\ref{figure:field}. The two electronics cabinets are mounted on each side of the mount.}
\label{figure:mount}
\end{figure}

We will point the six basic imagers using an Astelco NTM-500 mount used in an equatorial configuration. This mount is normally used in a German configuration, but has been modified by Astelco to have flanges on both sides. We have designed mechanical adaptors to allow us to mount three basic imagers on each flange. Figure~\ref{figure:mount} shows the design.

\subsection{Control Electronics}

The design of the control electronics has been adapted from the COATLI project\cite{watson-2016}. Each basic imager will be controlled by a MinnowBoard Max computer in an electronics cabinet. The MinnowBoard Max has a dual-core Intel Atom E3800 processor with 2 GB of RAM, 1000BASE-T ethernet, and two USB ports. One USB port will connect to the detector and one to the serial-to-USB adaptor for the focuser. The computers will run Ubuntu Linux. The MinnowBoards and power supplies will be mounted in two electronics cabinets, one on each side of the NTM-500 mount. Each cabinet will contain three MinnowBoards for the basic imagers and a fourth MinnowBoard for housekeeping tasks. The cabinets will have heating and ventilation to maintain them at a suitable temperature and humidity level. We will route 127 VAC power and ethernet cables through the NTM-500 mount to the electronics cabinets.

\subsection{Site}

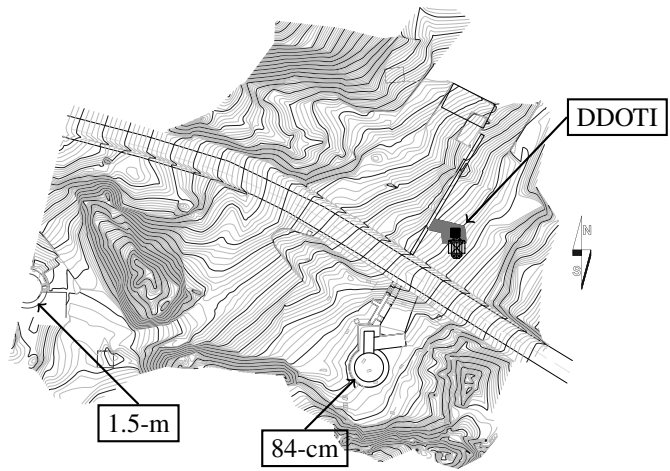
\begin{figure}
\begin{center}
\begin{tikzpicture}[scale=0.67]
\node at (0,0) {\includegraphics[width=10cm,angle=-13.5]{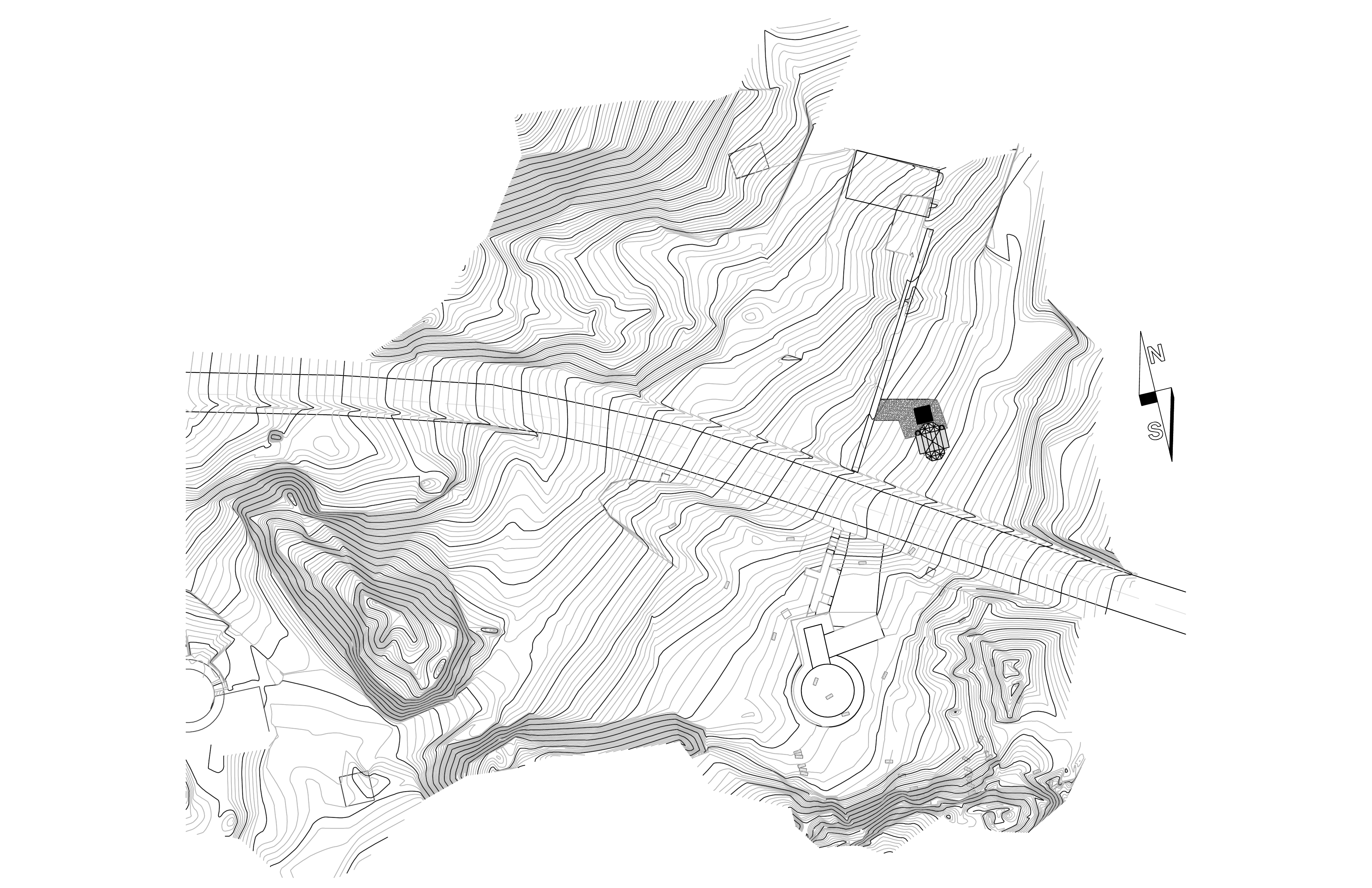}};
\draw[thick,->]  (-0.4,-4.2) node [anchor=north,rectangle,draw=black] {84-cm} -- (0.6,-3.2);
\draw[thick,->]  (-3.7,-3.7) node [anchor=north,rectangle,draw=black] {1.5-m} -- (-5.7,-1.7);
\draw[thick,->]  (4.8,2) node [anchor=west,rectangle,draw=black] {DDOTI} -- (2.8,-0.1);
\end{tikzpicture}
\end{center}
\caption{A map of the DDOTI site at the OAN/SPM.}
\label{figure:map}
\bigskip
\end{figure}

\begin{figure}
\begin{center}
\begin{tikzpicture}
\node at (0,0) {\includegraphics[width=10cm]{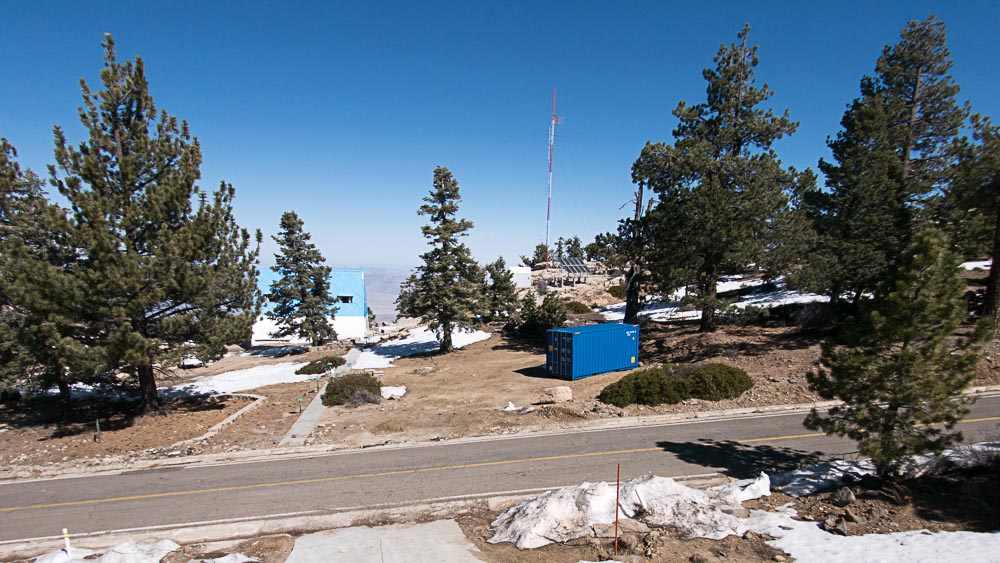}};
\draw[thick,->]  (-0.5,1.5) node [anchor=south,rectangle,draw=black] {DDOTI} -- (-0.5,-1.1);
\end{tikzpicture}
\end{center}
\caption{A photograph of the DDOTI site at the OAN/SPM taken from the balcony of the 84-cm telescope. The site is marked with an arrow. The blue shipping container will be removed prior to work on the site.}
\label{figure:site}
\bigskip
\end{figure}

Figures~\ref{figure:map}, \ref{figure:site}, \ref{figure:panorama} shows the site selected for DDOTI at $31^\circ 02^\prime 44.12^{\prime\prime}$ N $115^\circ 27^\prime 56.89^{\prime\prime}$ W. It is on flat ground about 50 meters to the north-north-east of the 84-cm telescope of the OAN/SPM.

\subsection{Enclosure, Platform, and Pier}

The design for the enclosure and platform has be adapted from the COATLI project\cite{watson-2016}.

We will use an Astelco ARTS station, which consists of a clamshell enclosure on a steel platform. This equipment is simple, robust, requires a minimum of maintainence, and has been proven in extreme conditions. The manufacturer guarantees it can open and close in a 90 km/h wind and survive a 180 km/h wind; in ten years of monitoring at the observatory the strongest gust we have seen is only 125 km/h.

\begin{figure}
\begin{center}
\includegraphics[width=\linewidth]{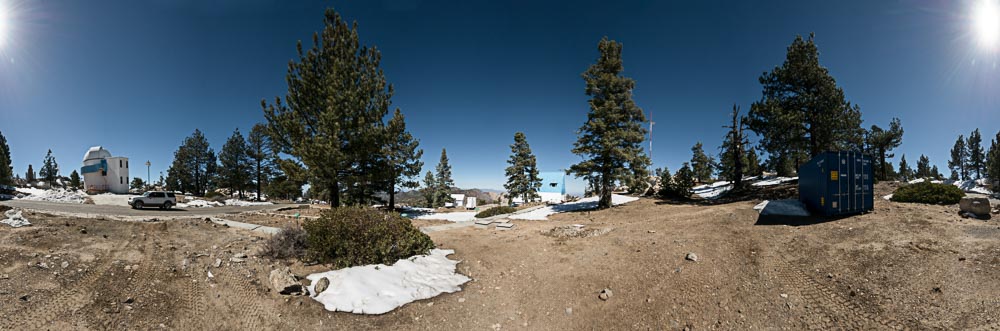}
\end{center}
\caption{A 360 {\deg} panoramic photograph taken from the DDOTI site at the OAN/SPM. The 84-cm telescope is the white dome in the left of the photograph. The three closest pine trees force us to elevate the mount to 6.5 meters to obtain access to zenith distances of 60 {\deg} at all azimuths.}
\label{figure:panorama}
\end{figure}

The platform places the rotation axis of the mount about 3.9 meters above the platform feet. A shown in Figure~\ref{figure:panorama}, there are three pine trees close to the site with would block access to the sky below zenith distances of 50 {\deg} if we used the steel platform alone. One option would be to remove these trees, but since the observatory is located in a national park this would require an environmental impact study which would add a delay of about nine months. Therefore, we will install the platform on 2.5 meter concrete columns to place the rotation axis of the mount about 6.4 meters above ground level. This will give access to the sky down to a zenith distance of 60 {\deg} in all directions.

In the standard Astelco ARTS station, the mount is attached to a vertical steel pier that is in turn attached to the platform. For DDOTI, this configuration will be modified. We will construct a 5-meter tall central column to serve as the telescope pier. This will then be extended by a further 1.4 meters with a bent steel pier whose final section will point to the northern celestial pole. This combination will give us an extremely solid telescope pier and will avoid mechanical interferences between the imagers and the steel pier.

\subsection{Control Software}

The control software will be adapted from the software we have used since 2011 to run the robotic Harold L. Johnson 1.5-meter telescope of the OAN/SPM and the RATIR instrument\cite{butler-2012,watson-2012}, with modifications from the COATLI project\cite{watson-2016} to control the Astelco NTM-500 mount and ARTS enclosure. One necessary modification will be to use the uncertainty maps distributed by the Fermi/GBM team to optimize our search strategy.

\subsection{Pipeline Software}

Again, our pipeline software will be adapted from the software we have used since 2011 for RATIR\cite{butler-2012}. Of course, it will require parallelization to handle approximately 288M pixels per minute in DDOTI rather than about 8M pixels per minute in RATIR. We will also modify the software to reject asteroids, known AGN, and flare stars, following iPTF\cite{singer-2015}. We will also increase the capacity of our existing data archive; DDOTI will generate about 1 TB of compressed data per month.

\section{Expected Performance}

\subsection{Image Quality}

Table~\ref{table:fwhm} shows the image quality budget, including the median seeing at SPM\cite{Skidmore:2009kw}, the intrinsic image quality of the RASA telescope, the small effect of having the detector window in an $f/2.2$ converging beam, and the field-dependent defocus caused by flexure. The final image quality is about 1.5 arcsec FWHM at the zenith and 1.8 to 2.7 arcsec FWHM at a zenith distance of 60 {\deg}. These images will be significantly undersampled by the 2 arcsec pixels of the ML50100 detector, so our effective image quality will likely be limited by sampling.

\begin{table}
\caption{Image Quality Budget at Zenith Distances of 0 and 60 {\deg}}
\label{table:fwhm}
\begin{center}
\begin{tabular}{lcc}
\hline
&$z = 0$ \deg&$z = 60$ \deg\\
&(arcsec)&(arcsec)\\
\hline
Seeing&0.78&1.18\\
RASA&1.10&1.10\\
Detector window&0.77&0.77\\
Flexure&0.00&0.0--2.0\\
\hline
Total&1.55&1.79--2.68\\
\hline
\end{tabular}
\end{center}
\end{table}

\subsection{Efficiency}

The RASA has Celestron's ``Starbright XLT'' coatings. Celestron states that these give a mirror reflectivity of 92.5\% and a corrector plate transmission of 97.4\% between 400 and 750 nm. The RASA additionally has four lenses and in its prime-focus corrector lens group. If we assume these each have the same transmission as the corrector plate, the total transmission of the RASA (excluding the window) is 81\%. We assume the RASA window is replaced by a filter with a transmission of 95\%. We assume the detector window has a transmission of 99\%. We assume the detector has a mean quantum efficiency of 55\%. The mean total efficiency of the system, excluding the atmosphere and the central obscuration, is 42\% in 400--750 nm.

The mean atmospheric extinction at 550 nm at SPM is about 0.15 magnitudes per airmass\cite{Schuster:2001to}. At a mean airmass of 1.5, this corresponds to an atmospheric transmission of 81\%. This reduces the mean effective efficiency of the system to 34\% in 400--700 nm, excluding the central obscuration.

We have no information on the efficiency of the RASA at longer wavelengths. It would not be surprising if the coatings degraded quickly beyond 750 nm. We will empirically determine the efficiency of the system as a function of wavelength by using a pupil mask to stop the telescope down to 12.5 mm and then observing bright standard stars with 50 nm filters from 350 to 1000 nm.

\subsection{Limiting Magnitude}

We estimate the zero point of the system, the count rate for a star with an AB magnitude of zero, to be $6.0 \times 10^8\ \mbox{electron}\ \mbox{s}^{-1}$ at 1.5 airmasses. We adopt sky brightnesses for SPM in dark/bright time of 21.9/19.2 $\mbox{mag}\ \persqarcsec$ in $g$, 21.1/19.6 $\mbox{mag}\ \persqarcsec$ in $r$, and 19.9/18.7 $\mbox{mag}\ \persqarcsec$ in $i$. With these, the sky rates are 11/50 $\mbox{electron}\ \mbox{s}^{-1}\ \mbox{pixel}^{-1}$. We assume a negligible dark current. We assume a read noise of 11 electrons.

To be definite, we assume a 400-800 nm bandpass and image quality of 4 {\arcsec} (2 pixels) FWHM. We use an aperture of 6.0 arcsec (3 pixels) diameter, which is close to optimal in the background-limited case, and assume a Gaussian profile, so 71\% of the light is within the aperture.

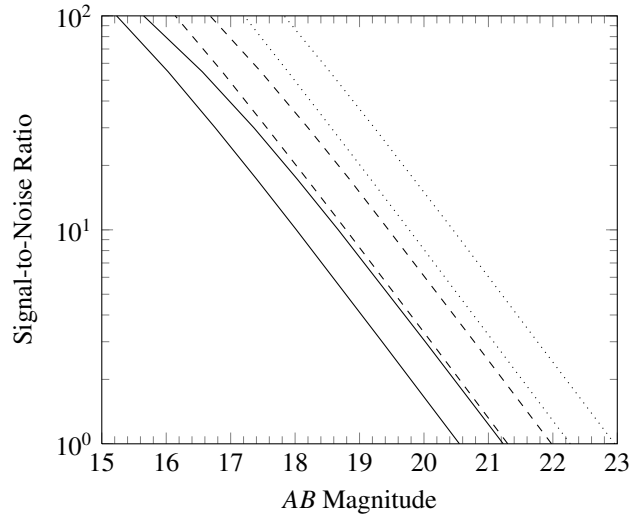
\begin{figure}
\begin{center}
\begin{tikzpicture}
\begin{axis}[
   xlabel={$AB$ Magnitude},
   ylabel={Signal-to-Noise Ratio},
   ymode=log,
   ymin=1,
   ymax=100,
   xmin=15,
   xmax=23,
   xtick={14,15,16,17,18,19,20,21,22,23,24},
   minor x tick num=4,
]
\addplot[black] table[x index=1,y index=0]{ddoti-limiting-magnitude.dat};
\addplot[black] table[x index=2,y index=0]{ddoti-limiting-magnitude.dat};
\addplot[black,dashed] table[x index=3,y index=0]{ddoti-limiting-magnitude.dat};
\addplot[black,dashed] table[x index=4,y index=0]{ddoti-limiting-magnitude.dat};
\addplot[black,dotted] table[x index=5,y index=0]{ddoti-limiting-magnitude.dat};
\addplot[black,dotted] table[x index=6,y index=0]{ddoti-limiting-magnitude.dat};
\end{axis}
\end{tikzpicture}
\end{center}
\caption{The limiting magnitude as a function of the SNR in $1 \times 60$ second (solid lines), $4 \times 60$ second (dashed lines), and $24 \times 60$ second (dotted lines) in bright (left) and dark (right) time.}
\label{figure:limiting-magnitude}
\end{figure}

Figure~\ref{figure:limiting-magnitude} shows the expected limiting magnitude for $1 \times 60$, $4 \times 60$, and $24 \times 60$ second exposures. A typical example is the $10\sigma$ limiting magnitude in 60 seconds, which is expected to be $r \approx 18.7$ in dark time and $r \approx 18.0$ in bright time. In $4 \times 60$ second exposures, the corresponding limits are magnitudes 18.8 and 19.5.

\section{Development}

\subsection{DDOTI Prototype}

We acquired a RASA, focuser, and ML16803 CCD for initial tests in Mexico City in February 2016. Obviously, the sky in the city is enormously bright, but we were able to demonstrate control of the CCD and focuser and measure the effect of flexure on the focal plane stability. We were not able to measure throughput (because of the heavy atmospheric extinction) or image quality (because the ML16803 has 9 {\micron} or 3 arcsec pixels).

We will install this prototype in parallel with the COATLI telescope\cite{watson-2016} at the OAN/SPM in September 2016. At this point, we will also substitute a ML50100 CCD with 6 {\micron} or 2 arcsec pixels. We will proceed to characterize the system, better determine the image quality, determine the efficiency as a function of wavelength, and determine if a filter is required to avoid chromatic aberration.

Once the definitive DDOTI/SPM instrument is installed, we will continue to use the prototype telescope for further projects. One is these is the DIRTI project, lead by Butler, which aims to implement a wide-field infrared imager.

\subsection{DDOTI/SPM}

We aim to install the complete six-telescope DDOTI at SPM in the spring of 2017. The project is fully funded. We aim to complete the civil work in fall 2016, but installation of the platform, enclosure, and mount will have to wait for better weather in spring 2017. We expect to be able to start observing almost immediately, since most of the software will be developed for the COATLI telescope and for the DDOTI prototype on the COATLI mount.

\subsection{DDOTI/OHP}

As we have mentioned, one of the aims of the DDOTI project is to install a network of similar telescopes around the world and thereby provide better sky coverage and earlier access to GRBs. We are investigating the possibility of locating the second DDOTI at either the Observatoire de Haute Provence (OHP) or the Observatoire de la Côte d’Azur at Calern. Both of these have relatively good weather, but there are concerns about light pollution at both. Neither have especially good seeing, but this is not important for DDOTI.

To move towards this goal, we aim to install a single basic imager in parallel with the IRiS telescope\cite{basa-2016} at OHP. The IRiS telescope is very similar to COATLI: it has a Astelco 50-cm telescope on an NTM-500 mount. This will allow us to better characterize OHP as a possible site for a full installation, especially the sky brightness, and will allow is to observe a subset of Fermi GRBs, perhaps those that can be observed in the first hour after burst.

\subsection{Beyond SPM and France}

We are actively pursuing funds and partners to install other DDOTI imagers around the world, to take advantage of the low unit cost of about US\$500,000 and high productivity.

\section*{Acknowledgements}

The DDOTI prototype and DDOTI/SPM has been funded by the Instituto de Astronomía of the UNAM, CONACyT proposals 260369, and 271117 (Laboratorio Nacional de Astrofísica en San Pedro Mártir), UNAM/PAPIIT project IG100414, and CRESST.

\bibliography{ddoti} 
\bibliographystyle{spiebib} 

\end{document}